\documentclass[aps,pra,twocolumn,groupedaddress]{revtex4-1}

\pdfoutput=1

\usepackage{graphics}
\usepackage{graphicx}
\usepackage{epsfig}
\usepackage{color}
\usepackage{amsmath}
\usepackage{bm}
\usepackage{overpic}
\usepackage[caption = false]{subfig}

\begin{document}




\title[Controlling quantum numbers and light emission of Rydberg states]{Controlling quantum numbers and light emission of Rydberg states via the laser pulse duration}

\author{L. Ortmann$^{1,2}$}
\email[]{ortmann.1@osu.edu}
\author{C. Hofmann$^{1,3}$}
\author{I. A. Ivanov$^{4}$}
\author{A. S. Landsman$^{1,2,5}$}
\email[]{landsman.7@osu.edu}
\affiliation{$^1$Max Planck Institute for the Physics of Complex Systems, N\"othnitzer Stra{\ss}e 38, D-01187 Dresden, Germany}
\affiliation{$^2$Department of Physics, Ohio State University, 191 West Woodruff Ave, Columbus, OH, 43210}
\affiliation{$^3$Department of Physics \& Astronomy, University College London, Gower Street, London WC1E 6BT, United Kingdom}
\affiliation{$^4$Centre for Relativistic Laser Science, Institute for Basic Science, Gwangju 61005, South Korea}
\affiliation{$^5$Department of Physics, Max Planck Postech, Pohang, Gyeongbuk 37673, Republic of Korea}

\date{\today}

\begin{abstract}
High Harmonic Generation (HHG) creates coherent high frequency radiation via the process of strong field ionization followed by recombination.  Recently, a complementary approach based on Frustrated Tunnel Ionization (FTI) was demonstrated \textit{(Nature Photonics \textbf{12}, 620 (2018))}.  It uses spectrally separated peaks created by lower quantum number Rydberg states to produce coherent extreme ultraviolet (EUV) light.  While much is understood about enhancing emission from HHG by controlling recombining electron trajectories, relatively little is known about controlling the quantum number distribution of Rydberg states.  This distribution is generally believed to be determined primarily by field strength and laser frequency.  We show that, in fact, it also changes significantly with the duration of the laser pulse: increasing pulse duration depletes lower-lying Rydberg states, thereby substantially decreasing EUV yield. Using electron trajectory analysis, we identify elastic recollision as the underlying cause.  Our results open a door to greater control over production of coherent high frequency radiation, by combining FTI and HHG mechanisms, and also improved the interpretation of molecular imaging experiments that rely on elastic electron recollision. 
\end{abstract}


\maketitle

Attosecond science investigates and makes use of phenomena that result from the interaction of a short-pulsed laser of extremely high intensity with atoms, molecules or solids \cite{krausz2009attosecond,scrinzi2005attosecond}. 
If the field strength of the laser field is comparable to the Coulomb force of an atom, the laser field can bend the Coulomb potential so strongly that a barrier is formed through which an electron can tunnel out of the atom \cite{keldysh1965ionization,corkum1993plasma,schafer1993above,ivanov2005anatomy}. After tunneling, the electron does not necessarily leave the atom for good, but can get captured in a Rydberg state, thus creating a neutral excited atom \cite{nubbemeyer2008strong,shvetsov2009capture, emmanouilidou2012multiple, mckenna2012frustrated}.

The relevance of this effect, referred to as `frustrated tunneling ionization' (FTI), already becomes clear by the observation that under typical strong field conditions 10-20\% of tunnel ionized electrons are trapped in Rydberg states \cite{nubbemeyer2008strong}, thus affecting many more electrons than other post-tunnel ionization processes such as high-harmonic generation (HHG) or double ionization by collision \cite{walker1994precision,eilzer2014steering}.
FTI not only explains the significant reduction of ionization rates \cite{nubbemeyer2008strong}, but can also be used to e.g.~calibrate laser intensities \cite{eichmann2013observing}, study nonadiabatic effects \cite{ortmann2018dependence}, probe the spatial gradient of the ponderomotive potential in a focused laser beam \cite{wells2004ionization}, or control the motion of neutral atoms in strong laser fields \cite{eichmann2009acceleration,eilzer2014steering}.

Rydberg states populate different quantum numbers, the distribution of which is important for the characterization of the excited neutral atoms, e.g.~in terms of their lifetime before decaying into metastable states \cite{nubbemeyer2008strong}. In the past decade, the distribution of principal quantum numbers has helped understand the stability of excited states under the influence of a second laser pulse \cite{eichmann2013observing,eilzer2014steering}, ionization channels and their closings \cite{li2014rydberg,zimmermann2017unified} as well as the effect of spatial gradients in the laser field \cite{zimmermann2018limit}. 

Recently, Yun et al. \cite{yun2018coherent} demonstrated a new source of coherent 
extreme-ultraviolet (EUV) light emission based on the recombination of Rydberg states produced by FTI, thereby complementing HHG and free induction decay after multiphoton excitation as coherent high frequency radiation sources \cite{beaulieu2016role,bengtsson2017space,beaulieu2017phase,chini2014coherent}. 
In this new radiation scheme, atoms are coherently excited into a Rydberg state $\Psi_R$ via FTI. In superposition with the ground state $\Psi_0(r)$, EUV light is emitted coherently whose frequency $\omega_{EUV}$ is determined by the difference between $I_p$, the energy of  $\Psi_0(r)$, and  $-0.5/n^2$, the energy of $\Psi_R(r)$:
\begin{equation}
	\omega_{EUV} = I_p - \frac{0.5}{n^2} \label{eq:EUV_freq}
\end{equation}
with the principal quantum number $n$. 
We use atomic units, unless stated otherwise.

Production of coherent EUV radiation via FTI relies on spectrally separated peaks created by low-lying (or small $n$) Rydberg states \cite{yun2018coherent}.  It is therefore important to find realistic laser parameters that preferentially populate these states.  Here, we show that
low-lying Rydberg states can be populated by using shorter, but experimentally realistic, pulses.  

\begin{figure}
	\includegraphics[width=\columnwidth]{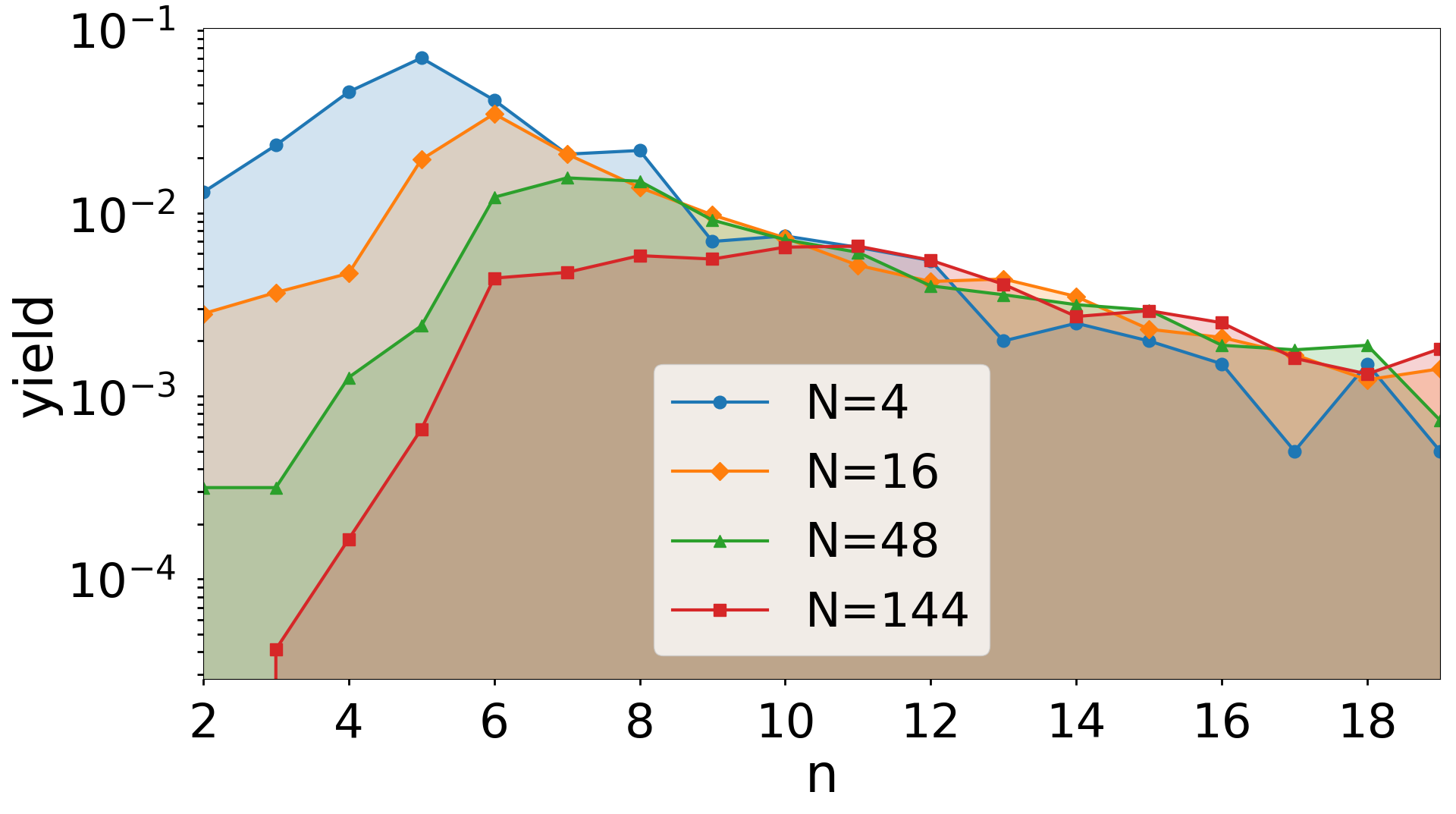}  
	\caption{Distribution of principal quantum numbers $n$ for various pulse durations specified by the number of cycles $N$. 
		Increasing the pulse duration diminishes the overall Rydberg yield and especially reduces the occupation of small principal quantum numbers. 
		The result was obtained from CTMC simulations with hydrogen at a laser intensity of $I=1.5 \times 10^{14} \mathrm{W/cm^2}$ and central wavelength $\lambda = 800 \mathrm{nm}$. }
	\label{fig:n_distro_depends_on_pulse_duration}
\end{figure}

Our results might seem surprising, as it is generally believed that quantum number populations are determined primarily by the peak field strength, $\mathcal{E}_0$, and laser wavelength, $\omega$, \cite{burenkov2010new,volkova2011ionization,fedorov2012interference,li2014fine,Potvliege1993,Morishita2013}, with the most likely principal quantum number given by \cite{nubbemeyer2008strong}
\begin{equation}
	\tilde n \approx \frac{\sqrt{\mathcal{E}_0}}{\omega}. 
	\label{eq:tilden}
\end{equation} 
In contrast, we find that increasing pulse duration (while keeping $\mathcal{E}_0$ and $\omega$ fixed) increases $\tilde n$, diminishes the overall Rydberg yield and dramatically reduces the occupation of small principal quantum numbers, which are necessary for the production of coherent EUV light via FTI  \cite{yun2018coherent}.  This is illustrated in Fig.~\ref{fig:n_distro_depends_on_pulse_duration}, which shows how low $n$ states are dramatically depopulated with increasing number of cycles, while high-$n$ states are relatively unaffected, resulting in a shift to higher $\tilde n$ values.  
We explain the underlying physical mechanism as arising from recollision with a parent ion, which occurs sooner for low-$n$ states. 
Our conclusions are illustrated with analysis of electron trajectories and supported by Classical Trajectory Monte Carlo (CTMC) simulations and the solution of the time-dependent Schr\"odinger equation (TDSE).
Note that the identified mechanism cannot be captured within the strong-field approximation framework, such as was recently used to investigate FTI as a source of coherent EUV radiation \cite{mun2018strong}.

Within the dipole approximation, a laser pulse linearly polarized along the z-direction is given by
\begin{equation}
	\vec{\mathcal{E}}(t) = \mathcal{E}_0 \cos(\omega t) \cos^2\left(\frac{\omega t}{2 N}\right) \Theta\left(|t|<\frac{\pi N}{\omega}\right) \vec{e}_z,  \label{eq:electricField}
\end{equation}
where $N$ denotes the total number of optical cycles and $\Theta$ is the heaviside step function. 
Although there are quantum effects in Rydberg states, such as channel closings or certain molecular effects, that can only be explained by invoking the time-dependent Schr\"odinger equation \cite{popruzhenko2017quantum, lv2016comparative, zimmermann2017unified, li2014fine}, semiclassical simulations have been found to be a powerful tool for understanding phenomena related to Rydberg atoms \cite{nubbemeyer2008strong,shvetsov2009capture,liu2012low,landsman2013rydberg,xiong2016correspondence}. Therefore, in addition to solving the time-dependent Schr\"{o}dinger equation (TDSE) \cite{yun2018coherent,ivanov2014evolution} to reproduce the EUV spectrum, we use a Classical Trajectory Monte Carlo (CTMC) method \cite{hofmann2013comparison,ortmann2018dependence} with ADK and TIPIS initial conditions \cite{pfeiffer2012attoclock} to analyze the underlying pulse duration effects.\\ 

Figure~\ref{fig:n_distro_depends_on_pulse_duration} gives the distribution of the principal quantum numbers from CTMC simulations for hydrogen. Details of this method are given in the appendix. We can see how the occupation of Rydberg states with lower principal quantum numbers decreases as the pulse duration increases.

\begin{figure*}
	\includegraphics[trim={0 0 8cm 0},clip,totalheight=3.7cm]{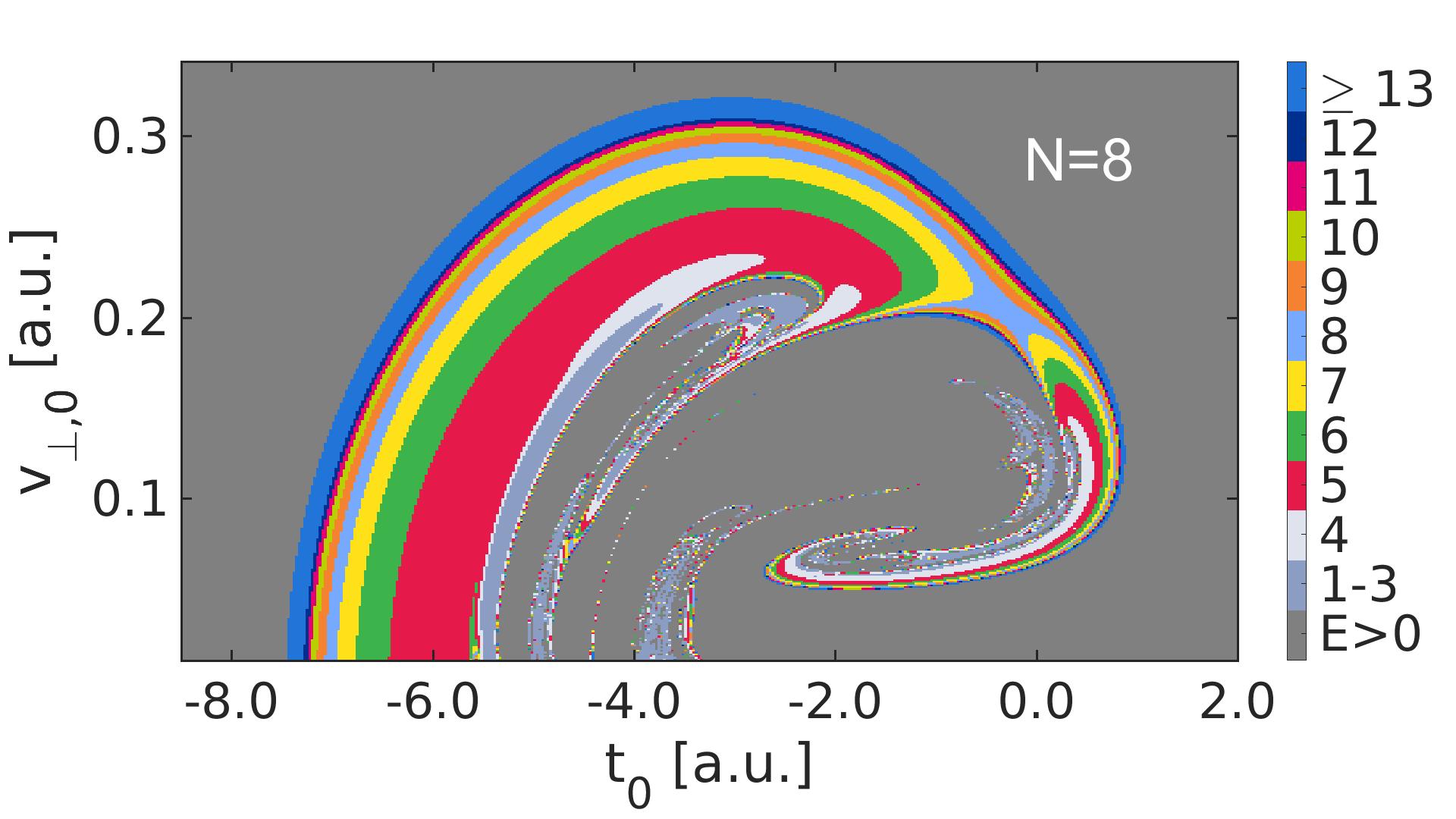}
	\includegraphics[trim={4.2cm 0 8cm 0},clip,totalheight=3.7cm]{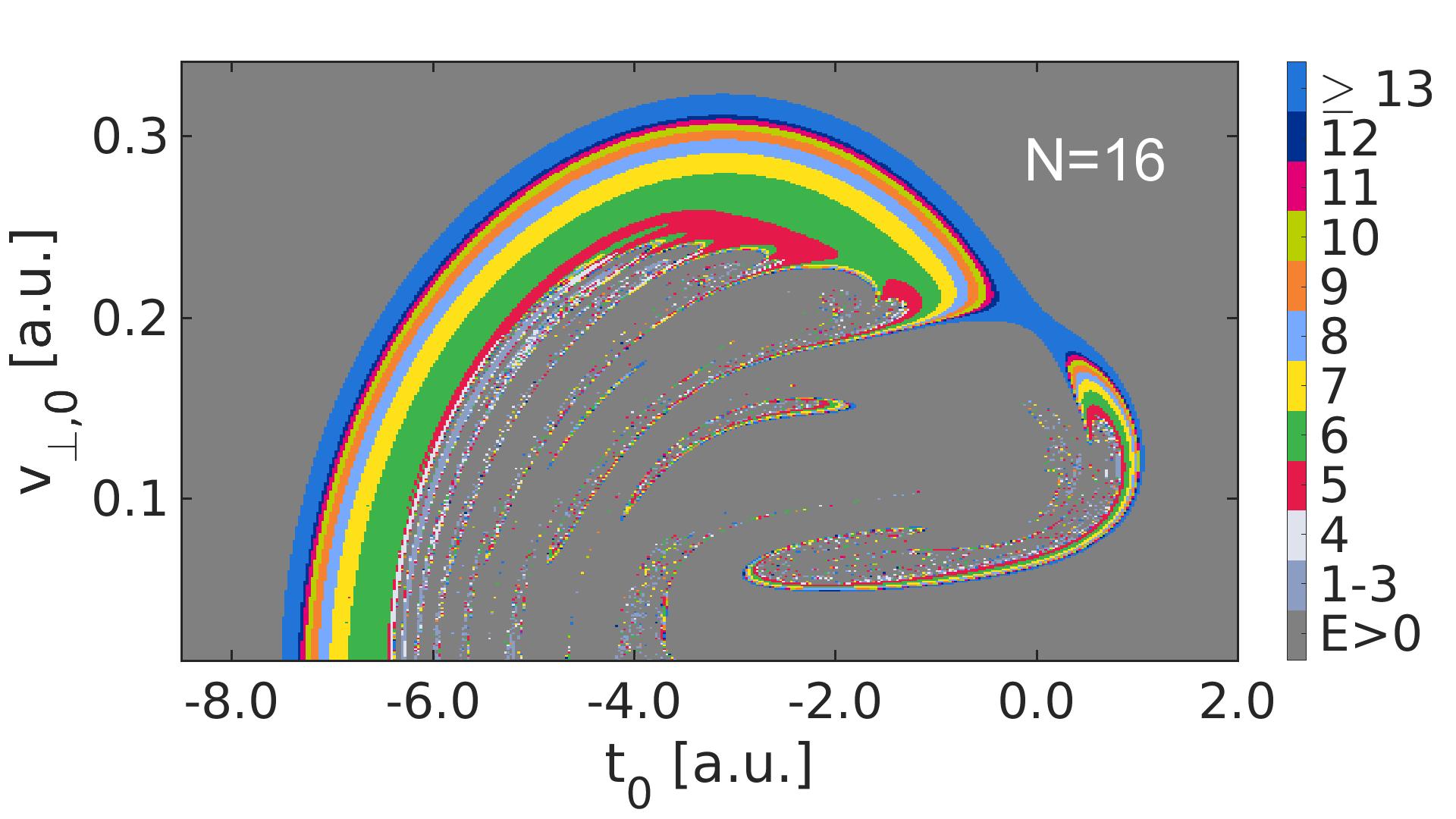}
	\includegraphics[trim={4.2cm 0 0cm 0},clip,totalheight=3.7cm]{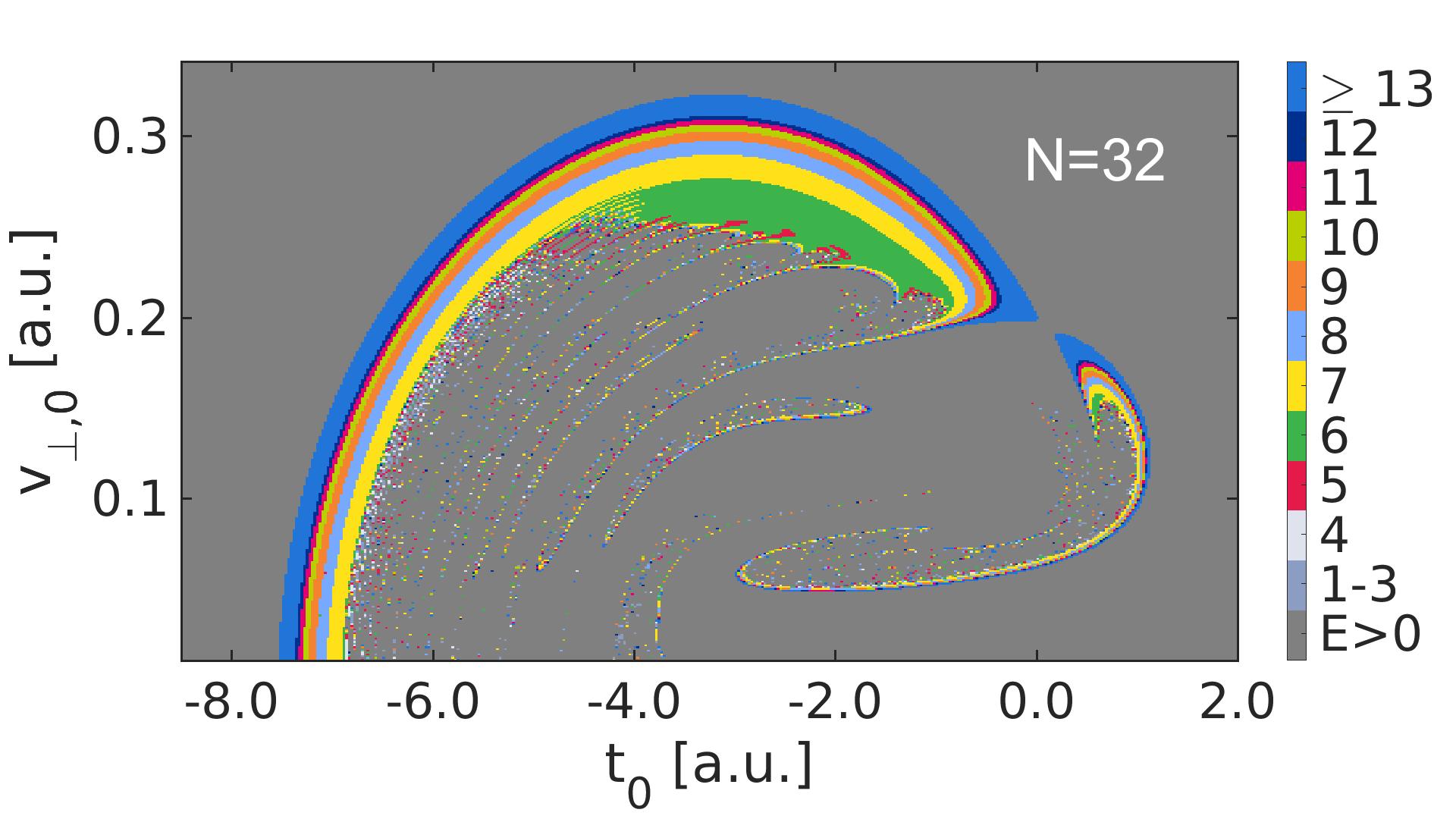}
	\caption{Map of the principal quantum number $n$ depending on the ionization time $t_0$ and the initial transverse velocity $v_{\perp,0}$. 
		The final energy from CTMC simulations was converted into the principal quantum number $n$ using $E = -0.5/n^2$ \cite{shvetsov2009capture,zhang2014generation}.
		Initial conditions which do not end up in a Rydberg state (positive total energy, $E>0$) are marked in gray. The pulse duration is given by the number of cycles $N$.}
	\label{fig:WaningMoon}
\end{figure*}

To understand this effect better, we look at trajectories ionized in the central half-cycle, where ionization is most likely. 
Figure~\ref{fig:WaningMoon} shows the dependence of the Rydberg yield, color-coded for different principal quantum numbers, on the ionization time $t_0$ (restricted to the central half-cycle) and the corresponding initial velocity $v_{\perp,0}$ of the electron. 
The Rydberg electrons populate a crescent-shaped area, with the smaller $n$ values located at the inner edge and high $n$-states populating the outer edge.  Note that most electrons originate before the peak of the laser pulse, corresponding to $t_0 < 0$, which is consistent with prior findings \cite{landsman2013rydberg}.
This nested crescent-shape quantum number distribution can be understood by considering that the earlier-born electrons are accelerated more by the laser field and have consequently higher energy after the pulse has passed, corresponding to higher-lying Rydberg states.

Figure~\ref{fig:WaningMoon} reveals that the value of $n$, as a function of the ionization time and initial velocity, does not change significantly with pulse duration. Rather, longer pulses deplete preferentially low-lying Rydberg states, corresponding to inner parts of the crescent.   
Note that this contrasts prior assumptions that the more loosely bound higher $n$ states are less likely to survive longer pulses \cite{shvetsov2009capture}.  

The successive depletion of low-$n$ Rydberg states with increasing pulse length leads to the shifting of the most likely principal quantum number, $\tilde n$, to successively higher values, as shown in Fig.~\ref{fig:n_distro_depends_on_pulse_duration}.  We find that this effect can be analytically described as follows,
\begin{equation}
	\tilde n = a+b \sqrt{N}
	\label{eq:sqrt}
\end{equation}
where $a$ and $b$ depend on laser frequency and field strength and $N$ is the number of laser cycles. An intuitive explanation for the linear dependence of $n$ on $\sqrt{N}$ is given in appendix \ref{appendix:sqrt_N_dependence_intuitive}.
Note that for short pulses, this result is compatible with the long-known Eq.~\eqref{eq:tilden} (see appendix \ref{appendix:sqrt_N_dependence_numerical} for details).


To test the robustness of our results, we replace the $cos^2$-term in Eq.~\eqref{eq:electricField} by a constant envelope, 
obtaining almost the same results as shown in Fig.~\ref{fig:WaningMoon} (see appendix), with significant discrepancies only for very small pulse durations in the regime of $N \leq 4$.  Hence, the occupation of low-lying Rydberg states depends strongly on the amount of time that the electron spends in the laser field rather than a particular pulse shape.

\begin{figure}
	\includegraphics[width=0.9\columnwidth]{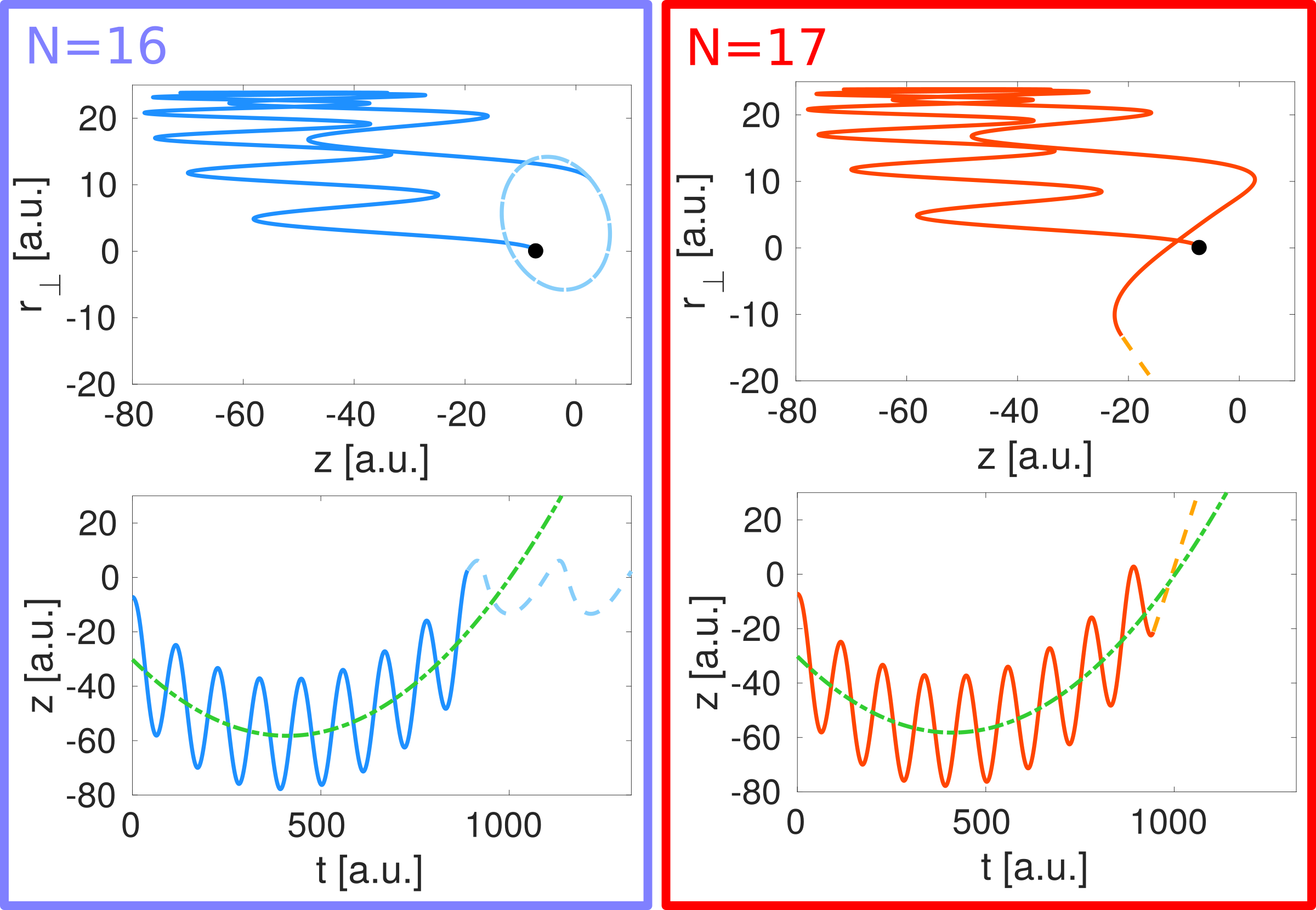}
	\includegraphics[trim={2.4cm 0 6.2cm 0},clip,width=0.9\columnwidth]{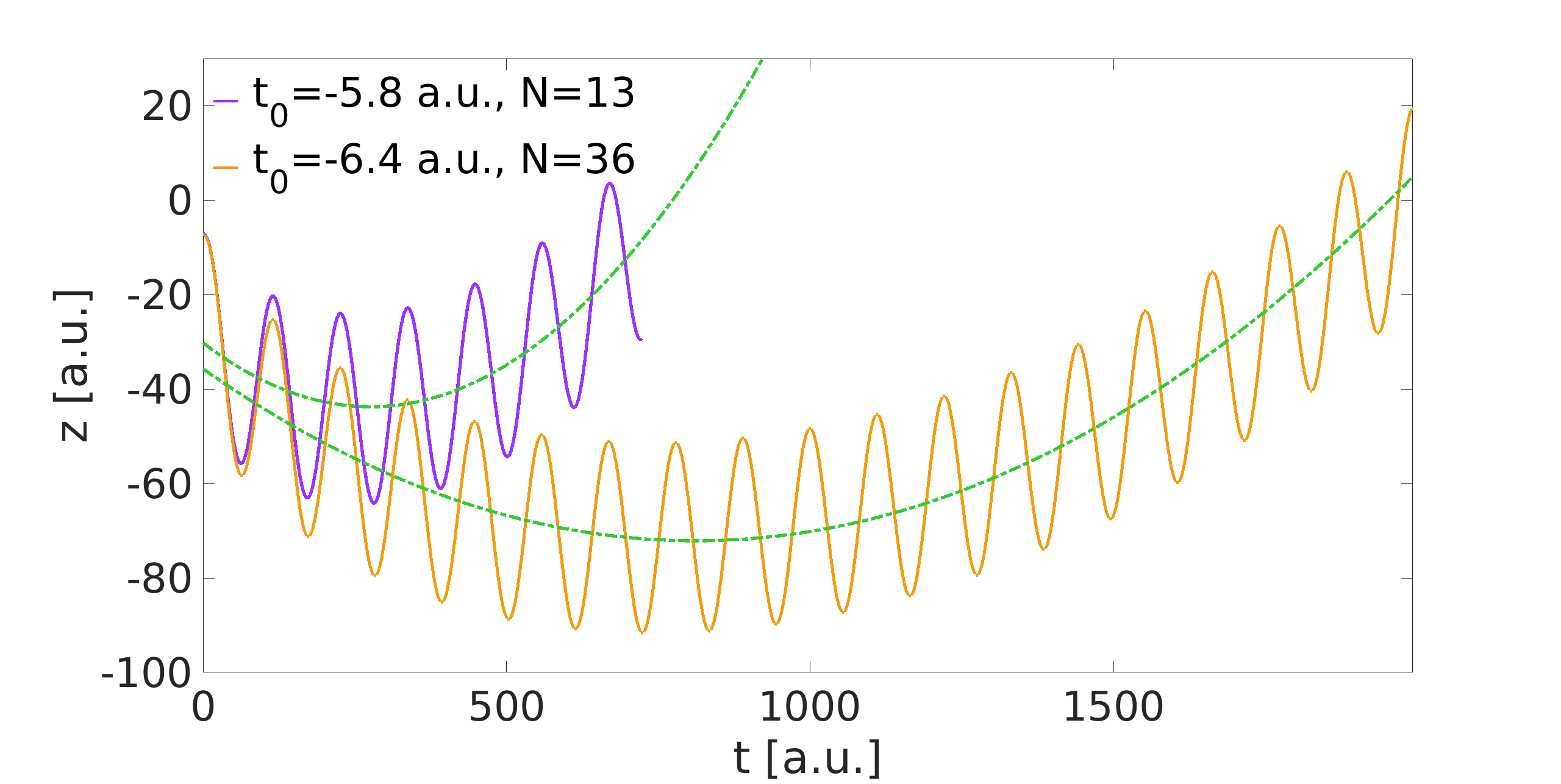}
	\caption{Trajectories released from a hydrogen atom in a laser field with $I=1.5\times10^{14}\mathrm{W/cm^2}$ and $\lambda=800 \mathrm{nm}$. Top panel: Trajectories released at $t_0 = -6.4~\mathrm{a.u.}$ and $v_{\perp,0}=0.09~\mathrm{a.u.}$ with a total number of cycles $N=16$ or $N=17$ in the pulse, respectively. $N=17$ exceeds the critical pulse duration leading to ionization. Bottom panel: Trajectories released with an initial transverse velocity $v_{\perp,0}=0.15~\mathrm{a.u.}$ for two different ionization times $t_0 = -5.8~\mathrm{a.u.}$ (purple/dark shorter trajectory) and $t_0 = -6.4~\mathrm{a.u.}$ (yellow/light longer trajectory) respectively, and correspondingly two different pulse durations, which are chosen such that a pulse duration of one more optical cycle would lead to the electron not ending up in a Rydberg state anymore. The green dashed-dotted lines show a parabolic fit through the value that $z$ oscillates around.
	}
	\label{fig:trajectories}
\end{figure}

To understand the physical mechanism that underpins the observed pulse length dependence, quantified in Eq.~\eqref{eq:sqrt}, we turn to analysis of individual electron trajectories.  
The top panels of Fig.~\ref{fig:trajectories} compare two electron trajectories with slightly different pulse durations -- one that still captures a particular electron into a Rydberg state ($N=16$) and another where the same electron is ionized ($N=17$). We can see that during the first optical cycles the electron is -- on cycle average -- driven away from the residual ion and comes closer to it later on, approximately following a concave-up curve, whose width and location of a minimum depends on the electron quantum number.  
This motion can be understood as resulting from competing effects of the attractive long-range Coulomb force of the ion and the laser field that drives the electron away from the ion.


The curves shown in Fig.~\ref{fig:trajectories}, depicting averaged motion of different electron trajectories,  illustrate how longer pulses lead to depletion of Rydberg states:  If a critical number of optical cycles (here $N=16$) is exceeded, the electron comes back to the ion, leading to elastic recollision and subsequent ionization. 
The recollision is elastic in the sense that the electron is only scattered from the Coulomb potential of the ion and does not exchange energy with the ion by e.g.~exciting or ionizing other electrons \cite{krausz2009attosecond}. Nonetheless, the electron can gain energy in this process due to the time-dependent electric field that is present during rescattering \cite{paulus1994rescattering, krausz2009attosecond}. This energy gain leads to the electron acquiring total positive energy, meaning it can leave the atom for good and thus ionization has taken place.

As the cycle-averaged electron motion is crucial in this process, rather than oscillation amplitude, which would be reduced by a $cos^2$-envelope, we can also understand why the
time spent interacting with the laser field, rather than a particular envelope shape, is important in this process. Overall, we find from CTMC simulations that the total Rydberg yield depends on the pulse duration rather than the particular shape of the pulse envelope (see appendix for details).



The bottom panel of Fig.~\ref{fig:trajectories} illustrates how higher-lying Rydberg states have significantly longer recollision times, thereby surviving longer pulses.
This panel shows two trajectories with the same initial velocity, but different ionization times, where an ionization time that is closer to the field maximum (at $t_0=0$) corresponds to a smaller principal quantum number. Fig.~\ref{fig:trajectories} shows that the electrons born near the field maximum
move along a narrower curve. This is mainly due to the smaller vector potential of the laser field closer to the field maximum. 
Therefore, the acceleration by the laser field driving the electron away from the ion is weaker and the Coulomb potential pulls the electron back faster. 
Consequently, the critical number of optical cycles above which the electron recollides and becomes ionized 
is smaller for the low-lying Rydberg electrons. 
This is the underlying reason behind the selective depletion of low-$n$ states with increasing pulse length.
Incidentally, a similar effect seems to play a role when a second laser pulse is used to probe the stability of Rydberg atoms \cite{eilzer2014steering,eichmann2013observing,stammer2020evidence}.  The use of the second pulse also preferentially depletes low principal quantum numbers by causing recollision and subsequent ionization of low-lying Rydberg states.


Note that the non-perturbative nature of the Coulomb potential is crucial to explaining the loss of low-lying Rydberg states with increasing pulse duration.  This is in contrast to other aspects of FTI, such as decline of Rydberg electrons with increasing ellipticity, which can be accurately described by neglecting the Coulomb potential altogether during the propagation \cite{landsman2013rydberg}.  
Another theoretical model \cite{shvetsov2009capture}, which neglects the Coulomb potential during propagation in the laser field, predicts a similar total yield but with a significantly different explanation, which relies on the assumption that it is the higher-lying Rydberg states that are preferentially ionized with increasing pulse duration.
Note that this stands in contrast to the results shown in Fig.~\ref{fig:WaningMoon}, where the outer boundary of the crescent (corresponding to high-$n$ Rydberg electrons) stays almost constant, 
while the Rydberg states vanish from the inner part of the crescent with increasing pulse duration. Thus, we see a counter-intuitive effect at work: The limit of $E=0$, corresponding to the loosely bound Rydberg electrons,  
is hardly affected by the pulse duration, whereas the deeply bound Rydberg states of more negative energies survive only short pulses.
Finally, existing quantum models, like the modified SFA in \cite{mun2018strong}, explain some features of the EUV radiation due to FTI \cite{yun2018coherent}, but neglect the Coulomb potential during propagation. Consequently these treatments cannot explain why shorter pulses are needed for the production of coherent EUV radiation via FTI, experimentally demonstrated in \cite{yun2018coherent}.  


\begin{figure}
	\includegraphics[width=\columnwidth]{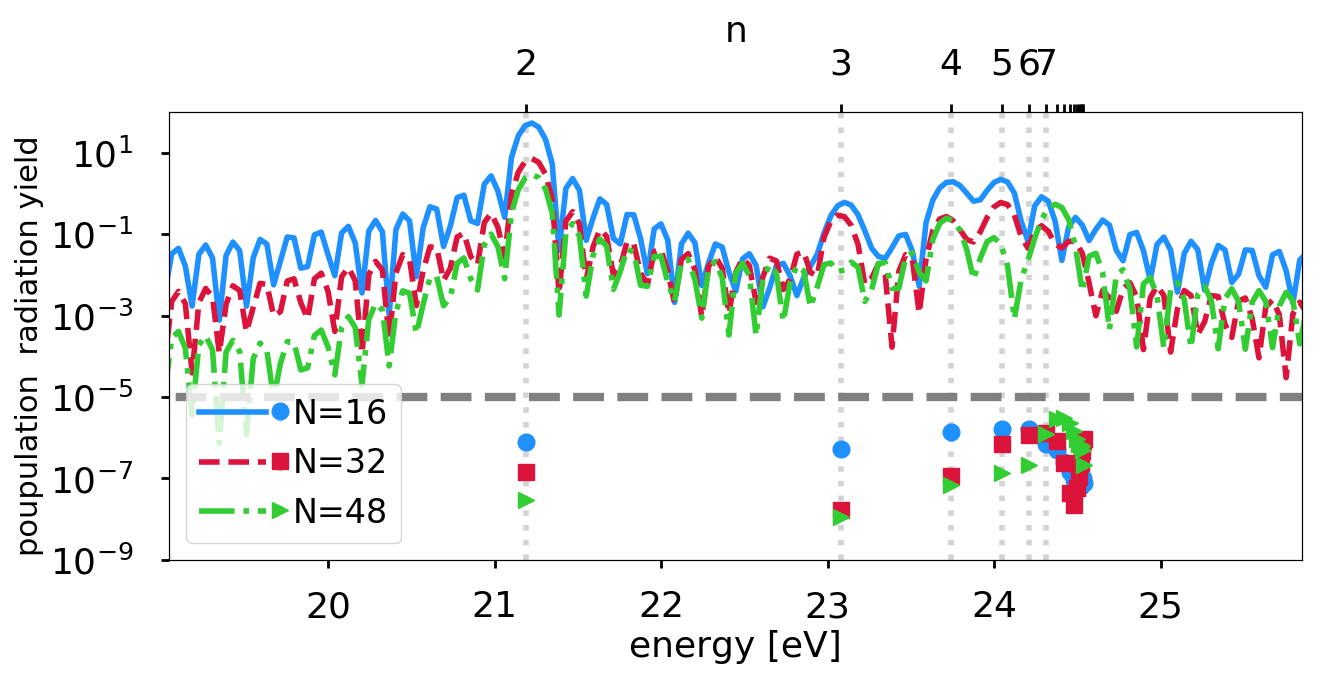}
	\caption{UV radiation yield (top panel) and occupation of quantum numbers (bottom panel) as a function of energy. The results were obtained in TDSE simulations using the parameters in \cite{yun2018coherent} for which the creation of EUV from FTI was experimentally realized: Ionization from helium at a laser intensity of $I=4.5 \times 10^{14}~\mathrm{W/cm^2}$ and a wavelength of $\lambda=730~\mathrm{nm}$.} 
	\label{fig:TDSE}
\end{figure}


To test our finding that longer pulses will suppress coherent EUV radiation due to depletion of lower-lying Rydberg states, we implement TDSE simulations, following the procedure in \cite{ivanov2014evolution}. 
The time evolution of the wave-function after the end of the laser pulse was obtained by first projecting the TDSE solution at the moment 
$t=T_1$, corresponding to the end of the laser pulse, on the subspace of bound states of the field-free
helium atom. Subsequent development of this projection in time can be described by the equation:
\begin{equation}
	\tilde\Psi(t)= \sum\limits_{k} \langle \phi_k|\Psi(T_1)\rangle \phi_k e^{-i\varepsilon_k(t-T_1)} \ ,
	\label{eq:de}
\end{equation}
where the sum on the right-hand side includes bound states of 
He atom with wave-functions $\phi_k$ and energies $\varepsilon_k$. 
Evolution equation \eqref{eq:de} was used to compute the
expectation value $\bar z(t)=\langle \tilde\Psi(t)|z|\tilde\Psi(t)\rangle$ of the dipole momentum operator, and
the radiation spectrum was obtained via Fourier transform of $\bar z(t)$.



Fig.~\ref{fig:TDSE} shows TDSE simulations of an EUV spectrum along with the occupation of principal quantum numbers.
As predicted, we see that the overall EUV yield decreases with increasing pulse duration. There is also a substantial depletion of lower-$n$ states (corresponding to $n \leq 5$) with increasing pulse duration, which is consistent with the single trajectory analysis above. The normalized distribution of the $n$-states, with $n$ on a linear scale for better visibility, is given in Fig.~\ref{pic:TDSE_vs_CTMC_n_distribution} in appendix \ref{appendix:CTMC_versus_TDSE} and shows how the maximum of the $n$-distribution shifts to larger values for longer pulses. CTMC simulations at the same laser and atomic parameters are given for comparison in this plot and show the same tendencies as the TDSE results.
These insights can be used to optimize coherent EUV light by employing shorter pulses or designing two-colour schemes that preferentially populate low-$n$ quantum states. 

\begin{figure}
	\includegraphics[width=\columnwidth]{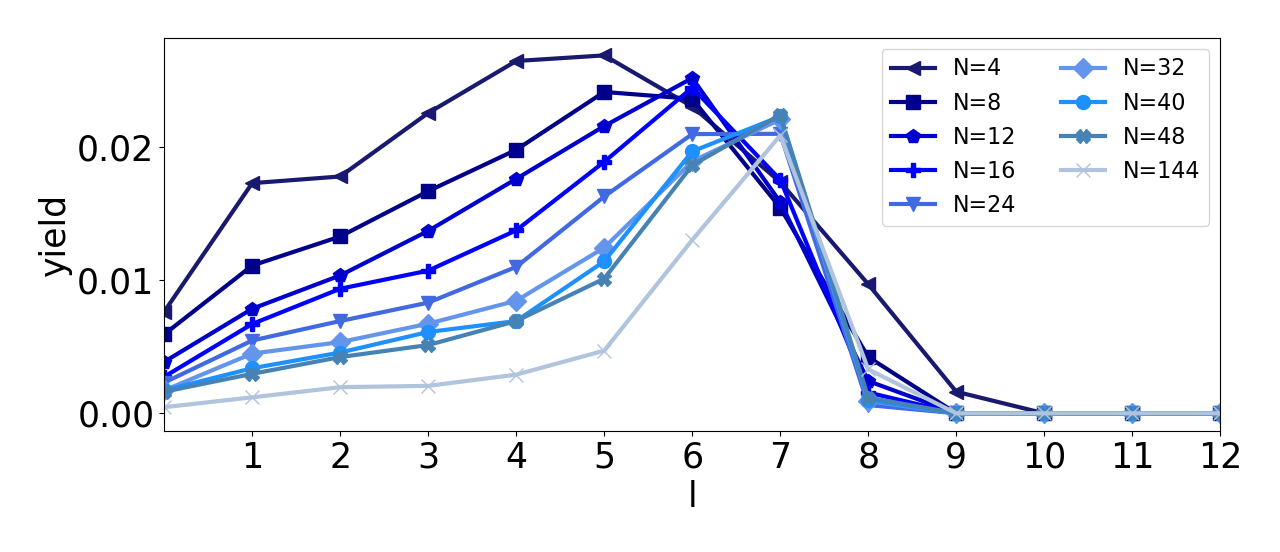}
	\caption{Yield of Rydberg states with angular quantum numbers $l$, obtained from semiclassical CTMC simulations with helium by calculating the classical angular momentum $L$ of the Rydberg states and obtaining $l$ from the relation $L^2=l \cdot (l+1)$ and rounding the resulting $l$ to an integer. The low $l$-states are depleted with increasing pulse duration.} 
	\label{fig:l-states}
\end{figure}

For the EUV emission from FTI \cite{yun2018coherent}, the relevance of occupation of low $n$-states is directly related to the occupation of low $l$-states, with the $p$-states being those that contribute most to the EUV radiation. As the angular quantum number $l$ can range from $0$ to $n-1$, we expect a depletion of low $n$-states to be accompanied by a depletion of low $l$-states, and thus also of the $p$-states ($l=1$) relevant for the EUV emission. That this is indeed the case can be seen in Fig.~\ref{fig:l-states}, where the angular quantum number $l$ of the Rydberg states obtained from the semiclassical CTMC simulations at the parameters of the experiment of Ref.~\cite{yun2018coherent} is depicted. It clearly shows that Rydberg states with $l < 6$, and thus also the p state, strongly decrease in their yield as the pulse duration is increased. This result is further supported by the $p$-state occupation obtained from the TDSE: Occupations of the states 2p and 6p, which were found in \cite{yun2018coherent} to be the  principal actors in  the EUV production, drop from $5.15 \times 10^{-6}$ to $6.83 \times 10^{-7}$ to $2.75 \times 10^{-7}$ (2p state for $N=16$, $N=32$, and $N=48$, respectively), and from $2.90 \times 10^{-7}$ to $2.47 \times 10^{-8}$ and $5.92 \times 10^{-8}$ (6p state for $N$=$16$, $32$, $48$).

To conclude, pulse duration complements other parameters known to affect the quantum number distribution, namely the intensity and wavelength of the field \cite{burenkov2010new,volkova2011ionization,fedorov2012interference,li2014fine,Morishita2013}. 
This leads to greater control over the new coherent EUV radiation scheme, where HHG peaks appear in the spectrum as well. As the HHG spectrum is highly sensitive to the intensity and wavelength of the incoming laser \cite{corkum1993plasma,schafer1993above,lewenstein1994theory}, pulse duration can serve as an independent knob controlling the relative contributions of FTI and HHG to the EUV spectrum.  This raises a possibility of enhancing attosecond pulses by combining FTI and HHG contributions.    

We identify elastic recollision as a key mechanism behind depletion of lower-$n$ Rydberg states with changing pulse duration.  
Moreover, it is possible to rescatter electrons at a well-defined time and energy determined by pulse duration.  Direct timing information is provided by the fact that for each electron trajectory, there is a critical pulse duration corresponding to recollision time, as was shown in Fig. \ref{fig:trajectories}. Normally time has to be inferred from some other observable, like HHG \cite{shafir2012resolving, fiess2011attosecond} or electron momenta distributions \cite{landsman2014ultrafast}.  
This provides a new, complementary tool for exploring recollision physics and an opportunity to test existing theoretical models.
This is particularly important for lower energy electrons, where the classical assumptions underlying the three-step model \cite{corkum1993plasma,schafer1993above} can break down. Improved  understanding of elastic recollision should lead to more accurate time-resolved molecular imaging, which relies on the analysis of electron momenta distributions \cite{walt2017dynamics,bian2012attosecond,meckel2014signatures}.

\begin{acknowledgments}
L.O. acknowledges support by the Max Planck Society via the IMPRS for Dynamical Processes in Atom, Molecules and Solids, and was supported in part by the Center for Emergent Materials, an NSF MRSEC, under Grant No. DMR-2011876.
C.H. acknowledges support by a Swiss National Science Foundation mobility fellowship.
I.A.I. acknowledges support from the Institute for Basic Science,  grant number IBS-R012-D1.
A.S.L. received supported by the Max Planck Center for Attosecond Science (MPC-AS).
\end{acknowledgments}

\appendix
\renewcommand\thefigure{\thesection\arabic{figure}}

\section{Description of the Classical-Trajectory Monte Carlo (CTMC) simulations}
In our Classical-Trajectory Monte Carlo (CTMC) simulations an ensemble of trajectories is started with the initial conditions chosen such that, for the whole ensemble, they follow the ADK probability distribution \cite{delone1991energy, ammosov1986tunnel} 
\begin{equation}
	P(t_0,v_\perp) \propto  \exp\left( - \frac{2 (2 I_p)^{3/2}}{3 \mathcal{E}(t_0)} \right)  \cdot \exp\left( -\frac{\sqrt{2 I_p}}{\mathcal{E}(t_0)} v^2_{\perp,0}  \right),
\end{equation}
with $t_0$ denoting the time of the electron's appearance at the tunnel exit and $v_{\perp,0}$ being its initial velocity perpendicular to the laser polarization. For a detailed description of the method we refer to \cite{hofmann2013comparison}. 
The coordinates of the tunnel exit are calculated on the basis of the energy conservation law in parabolic coordinates \cite{pfeiffer2012attoclock,fu2001classical,landau2013quantum}.
The electron trajectories are then propagated in the superposed potential of the laser and Coulomb field solving Newton's equations:
\begin{equation}
	\ddot{\vec{r}} = -\vec{\mathcal{E}}(t) - \frac{\vec{r}}{(\vec{r}^2 + a^2 )^{3/2}}.
\end{equation}
with soft core parameter $a^2 = 0.01$.

\section{Rydberg area with $\cos^2$ and with constant envelope}
\begin{figure*}
	\includegraphics[trim={0 0 8cm 0},clip,totalheight=3.7cm]{nMatrix_myColormap_I1comma50eplus14_lam8comma00eminus07_N_8.jpg}
	\includegraphics[trim={4.2cm 0 8cm 0},clip,totalheight=3.7cm]{nMatrix_myColormap_I1comma50eplus14_lam8comma00eminus07_N_16.jpg}
	\includegraphics[trim={4.2cm 0 0cm 0},clip,totalheight=3.7cm]{nMatrix_myColormap_I1comma50eplus14_lam8comma00eminus07_N_32.jpg}\\
	\includegraphics[trim={0 0 8cm 0},clip,totalheight=3.7cm]{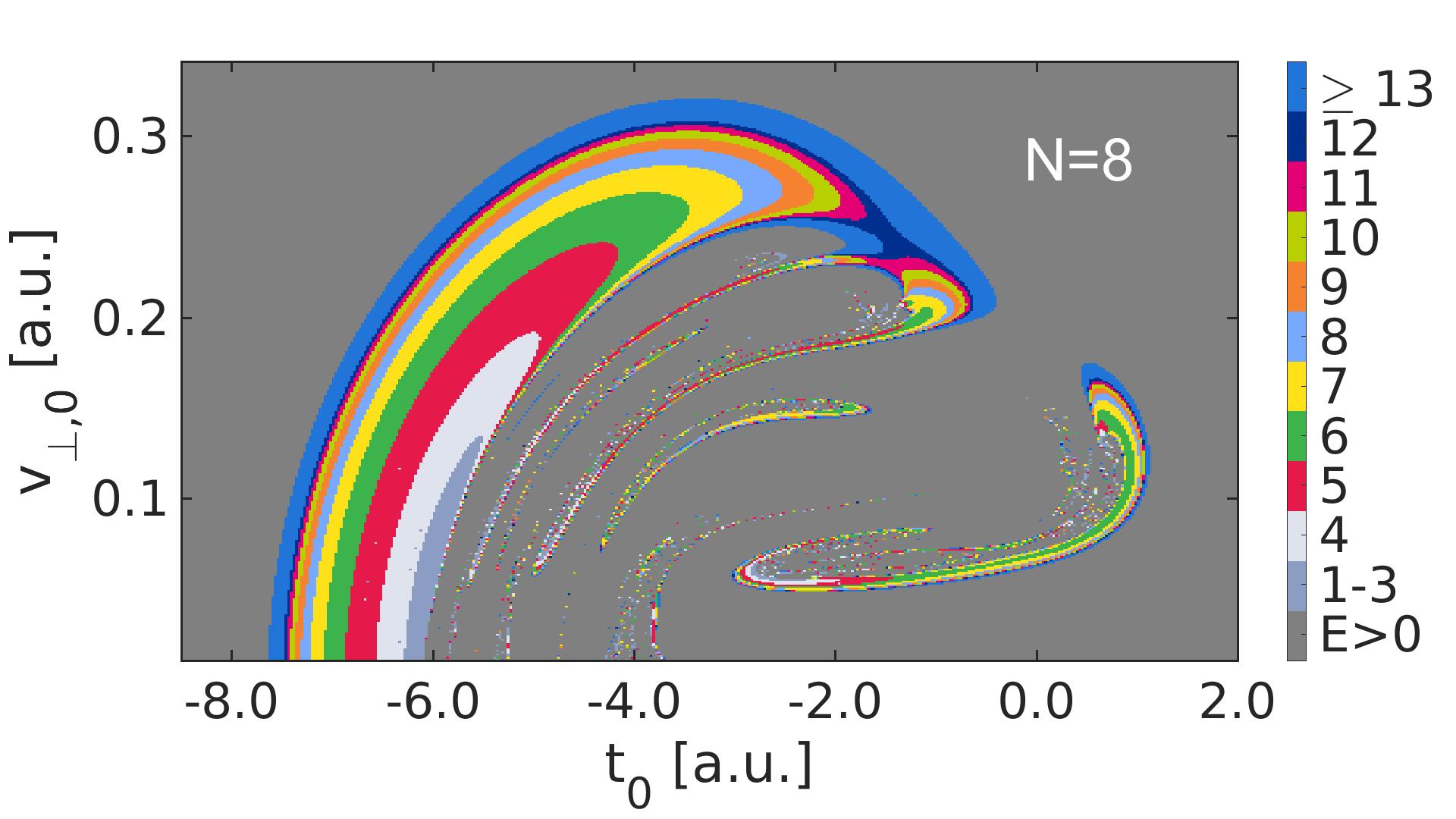}
	\includegraphics[trim={4.2cm 0 8cm 0},clip,totalheight=3.7cm]{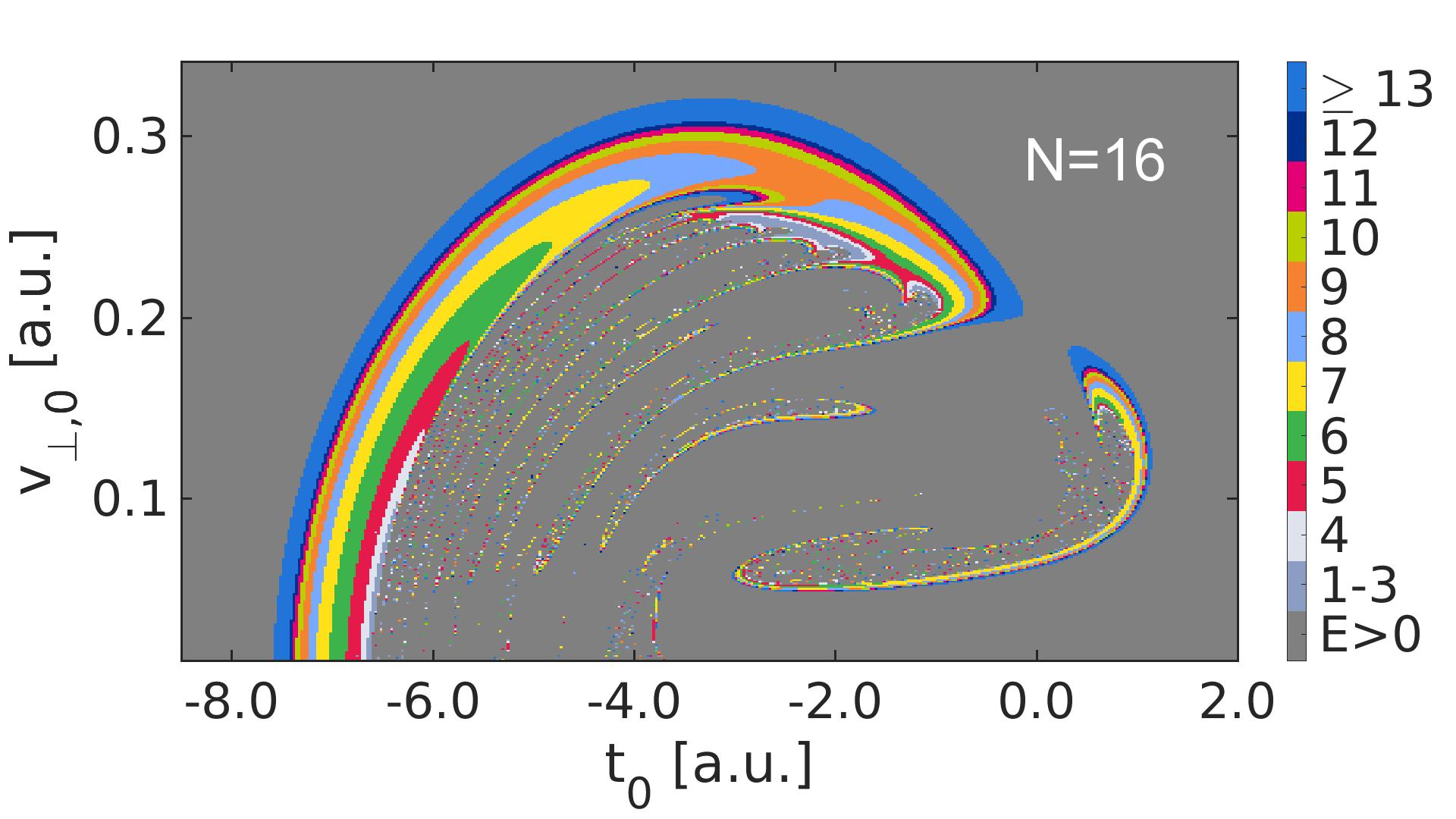}
	\includegraphics[trim={4.2cm 0 0cm 0},clip,totalheight=3.7cm]{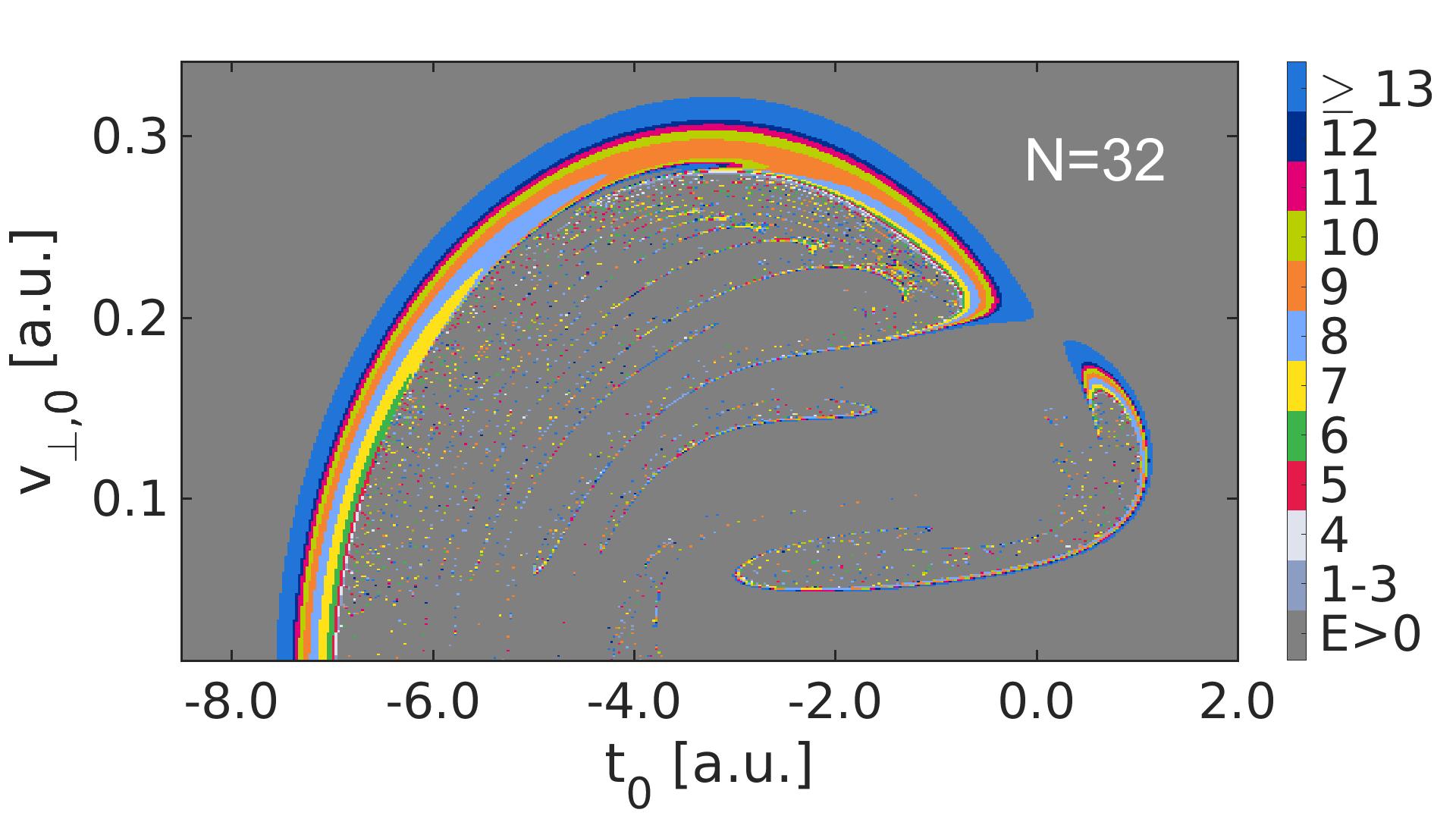}
	\caption{Top row: Pulse with $\cos^2$-envelope. This figure is identical to Fig.~\ref{fig:WaningMoon} in the main text and is reprinted here merely for easier comparison. Bottom row: Same as top row except for the constant envelope of the pulse used here.}
	\label{fig:WaningMoon_appendix}
\end{figure*}

To conclusively establish whether the loss of Rydberg electrons for longer pulses 
is caused by the longer time the electron spends in the laser field or by the different shapes of the envelope that the different pulse durations imply, additional simulations are performed.
Replacing the $\cos^2$-term in eq.~\eqref{eq:electricField} by a constant
we get almost the same crescents as the ones obtained using the full envelope. This can be seen from comparing the top and bottom row of Fig.~\ref{fig:WaningMoon_appendix}, with the former using a $cos^2$-envelope and the latter a constant pulse envelope. Significant discrepancies due to different types of envelopes are only observed for very short pulse durations in the regime of $N \leq 4$. 

We observe the same pulse duration effects independent of the envelope shape: In both the top and bottom row of Fig.~\ref{fig:WaningMoon_appendix}, increasing the pulse duration leads to a depletion of the `inner edge' of the crescent-shaped area, and in both cases the position of the quantum numbers are hardly affected by the pulse duration.
This shows that the main reason for the loss of low quantum number Rydberg states with longer pulse durations is the long time the electrons spend in the laser field rather than a particular envelope shape. 

\section{Dependence of the peak of the quantum number distribution on the pulse duration - Numerical aspects} \label{appendix:sqrt_N_dependence_numerical}

\begin{figure}
	\centering
	\includegraphics[width=\columnwidth]{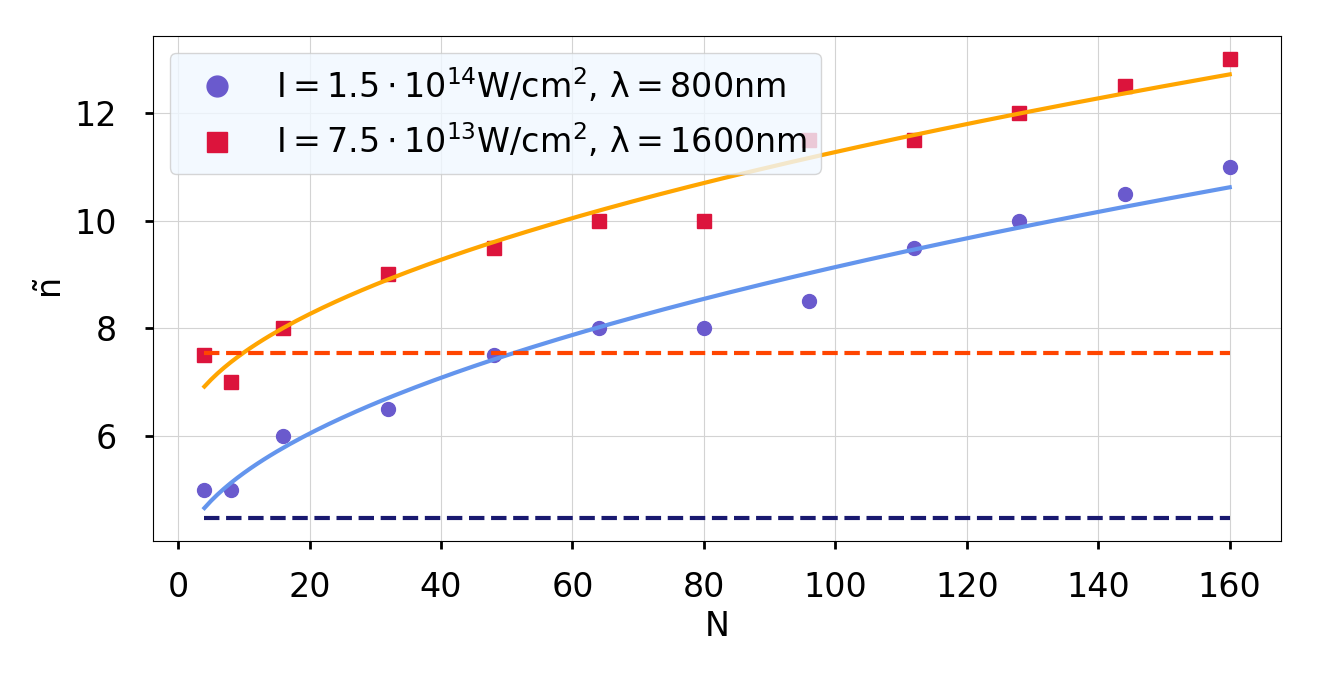} 
	\caption{Quantum numbers $\tilde{n}$ at which the quantum number distribution peaks as a function of the total number of optical cycles of the pulse for two different laser paramters specified in the legend (atomic target is hydrogen in both cases). Eq.~\eqref{eq:Fitting} is fitted to the data and shown as solid lines. The horizontal dashed lines are positioned at $\sqrt{\mathcal{E}}/\omega$ and approximately coincide with the peak position $\tilde {n}$ at the shortest pulse duration examined ($N=4$).}
	\label{pic:Jing_Su_1D_Gaussian_model_potential_problem_at_low_energies} 
\end{figure}

The successive depletion of low-n Rydberg states with increasing pulse length leads to the shifting of the most
likely principal quantum number, $\tilde {n}$ , to successively higher values, as shown in Fig.~\ref{fig:n_distro_depends_on_pulse_duration}. We find that this
affect can be analytically described by
\begin{equation}
	\tilde {n} = a + b\sqrt{N} \label{eq:Fitting}
\end{equation}
where $a$ and $b$ are fitting parameters whose physical meaning will become clear in the following. A depiction of this function can be found in  Fig.~\ref{pic:Jing_Su_1D_Gaussian_model_potential_problem_at_low_energies} and it shows that the different laser paramters chosen only affect significantly the offset $a$ but not the increase $b\sqrt{N}$. This is also confirmed numerically by comparing the fitting parameters b: We have $b=0.56$ for the blue curve and $b=0.54$ for the orange curve, which correspond to simulations at two different set of laser parameters (see legend). 

As should also become clear from Fig.~\ref{pic:Jing_Su_1D_Gaussian_model_potential_problem_at_low_energies}, the offset $a$ is directly related to the long-known estimate of the peak quantum number $\tilde {n}$ that does not account for pulse duration effects that was given in eq.~\eqref{eq:tilden} and which we reprint for quick reference
\begin{equation}
	\tilde {n} = \frac{\sqrt{\mathcal{E}}}{\omega}. \label{eq:simple}
\end{equation}
This value is depicted as dashed horizontal lines in Fig.~\ref{pic:Jing_Su_1D_Gaussian_model_potential_problem_at_low_energies} and we can see that it coincides with eq.~\eqref{eq:Fitting} for short pulses, here $N=4$. Thus it becomes clear that the long-known estimate for $\tilde {n}$ given by eq.~\eqref{eq:simple} is true only for short pulses in the few-cycle regime but that the peak position will shift due to depletion of low principal quantum numbers and that this shift can be described by the $b\sqrt{N}$ part of eq.~\eqref{eq:Fitting}.

A slightly more mathematical description is given in the following.
The offset $a$ is directly related to $\sqrt{\mathcal{E}}/\omega$, but it is not identical to it because $a$ is the value for $N=0$, which is an unphysically short pulse. Rather we find that the value of $\sqrt{\mathcal{E}}/\omega$ is attained for short pulses of about $N=4$ and we can write
\begin{equation}
	\sqrt{\mathcal{E}}/\omega = a + b\sqrt{4}.
\end{equation}
Since the parameter $b$ (fitting parameter) is found to be almost identical for the two strongly different laser parameters used in Fig.~\ref{pic:Jing_Su_1D_Gaussian_model_potential_problem_at_low_energies} we can use this value of $b \approx 0.55$ to give an estimate for the relation between the offset $a$ and the value $\sqrt{\mathcal{E}}/\omega$ from eq.~\eqref{eq:simple}
\begin{equation}
	\sqrt{\mathcal{E}}/\omega \approx a + 0.55\sqrt{4} = a + 1.1.
\end{equation}

For an intuitive physical explanation for the linear dependence of $n$ on $\sqrt{N}$ we refer to section \ref{appendix:sqrt_N_dependence_intuitive} of the appendix.

\section{Dependence of the peak of the quantum number distribution on the pulse duration - Intuitive explanation} \label{appendix:sqrt_N_dependence_intuitive}

We can intuitively understand the result of eq.~\eqref{eq:sqrt}, which tells us that the most likely principal quantum number  
$\tilde{n}$ grows as $\sqrt{N}$, by invoking that the mean distance of the Rydberg electron $r_n$ follows the dependence $r_n \simeq n^2$ \cite{volkova2011ionization}. As we have seen in Fig.~\ref{fig:trajectories}, the electron mean position follows a trajectory which can be approximately described by
\begin{equation}
	z_{mean}(t) = d \cdot (e-t)^2 + f \label{eq:parabolaRebuttal},
\end{equation}
with parameters $d$, $e$, and $f$.
Here, we restrict ourselves to the electron's motion along the z-direction, the polarization direction of the laser.
The extremum of the parabola is found at time $e$ and we can assume that it is attained after half of the propagation time: $e \approx N/2$, where $N$ is the number of cycles the electron propagates before it recollides with the ion. The parameter $f$ is the mean distance attained at this time $t=e$, i.e. $f$ is the largest mean distance from the ion.
Directly after ionization, at $t=0$, eq.~\eqref{eq:parabolaRebuttal} becomes
\begin{equation}
	z_{mean}(0) = d \cdot e^2 + f; \quad z'_{mean}(0) = -2de 
\end{equation}
and thus
\begin{eqnarray}
	f 	&=& z_{mean}(0) - d \cdot e^2 \nonumber\\
		&=& z_{mean}(0) + \frac{z'_{mean}(0)}{2} e \nonumber\\
		&\approx& z_{mean}(0) + \frac{z'_{mean}(0)}{2} \frac{N}{2}.
\end{eqnarray}
We can see that the maximum mean elongation $f$ scales linearly with $e \propto N$, suggesting that the overall mean elongation $r_n \simeq n^2$ grows linearly with $N$ and consequently $n \propto \sqrt{N}$, in agreement with the results obtained by numerical fits in eq.~\eqref{eq:sqrt}.

These considerations also show that the parabola's curvature radius and the electron's mean elongation will be larger if $|z'_{mean}(0)|$ is larger. As we can approximate the mean velocity offset $|z'_{mean}(0)|$ directly after ionization by $\mathcal{E}_0/\omega \cdot \sin(\omega t_0) \approx \mathcal{E}_0 t_0$, we can see clearly that an earlier birth, i.e. $t_0$ is more negative, leads to a larger $|z'_{mean}(0)|\approx |\mathcal{E}_0 t_0|$. As this results in a larger overall mean elongation, the earlier ionization time is related to a larger $n$, just as we have seen in Fig.~\ref{fig:WaningMoon}. 

\section{Comparison of TDSE and CTMC distribution of principal quantum numbers n} \label{appendix:CTMC_versus_TDSE}

\begin{figure}
	\centering
	\includegraphics[width=\columnwidth]{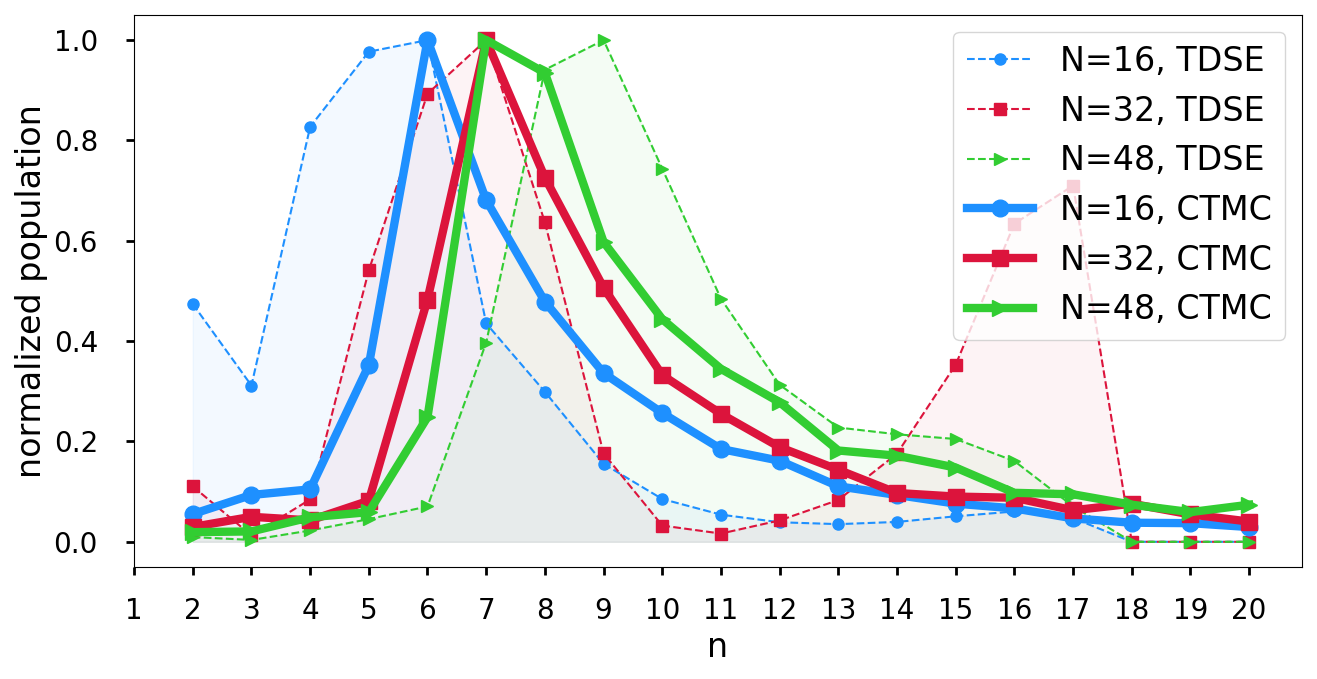} 
	\caption{Population of Rydberg states with principal quantum numbers $n$, normalized to the maximum value for better comparison. The round dots connected by dashed lines that enclose a lightly colored area represent the TDSE data, whereas the large square dots connected by thick solid lines represent the CTMC data. We can see that the maxima are found at equal or similar values for TDSE and CTMC for the 3 different pulse durations shown.}
	\label{pic:TDSE_vs_CTMC_n_distribution} 
\end{figure}

In addition to Fig.~4, which showed the EUV radiation spectrum obtained from TDSE (upper part of figure) and the absolute populations of the Rydberg states with principal quantum number $n$ (lower part of figure), we directly compare the relative $n$ distribution in CTMC and TDSE. The result is found in Fig.~\ref{pic:TDSE_vs_CTMC_n_distribution}. The laser and atomic parameters are the same as the ones used for the TDSE result in Fig.~\ref{fig:TDSE} and for the CTMC results in Fig.~\ref{fig:l-states}.

Fig.~\ref{pic:TDSE_vs_CTMC_n_distribution} shows that the principal quantum number distribution peaks at identical or similar values in the CTMC and TDSE result, and that this peak shifts to larger $n$ by about the same value in both methods as the pulse duration increases.

The increasing relative population at larger $n$ in the TDSE result for a pulse with $N=32$ cycles is interpreted as being due to interference effects that cannot be captured in the semiclassical CTMC simulations.

\bibliography{Ryd_lib}

%
%
%
%

\end{document}